\documentclass[aps,prb,reprint,groupedaddress]{revtex4-2}

\usepackage{graphicx}
\usepackage{amsmath,amsfonts,amssymb}
\usepackage{dcolumn}
\usepackage{bm}
\usepackage{braket}

\usepackage{enumitem}

\usepackage[T1]{fontenc}

\usepackage[breaklinks=true,colorlinks,allcolors=blue]{hyperref}

\begin{document}

\preprint{APS/123-QED}

\title{Simulation of temperature-dependent quantum gates in silicon quantum dots with frequency shifts}


\author{Yudai Sato}
  \email{4324529@ed.tus.ac.jp}
\author{Takayuki Kawahara}
  \email{kawahara@ee.kagu.tus.ac.jp}
\affiliation{ Department of Electric Engineering, Tokyo University of Science, Katsushika, Tokyo, 125-8585, Japan}


\date{\today}

\begin{abstract}
  To achieve quantum computing using semiconductor spin qubits, the spin qubits must be precisely controlled. However, unexpected noise limits this precision and prevents the implementation of error correction codes. Specifically, frequency shifts have been found to suppress one-qubit gate fidelity. Although the exact source of the frequency shifts remains unclear, several experiments have indicated a relationship between the frequency shift and heating from the control signal. Based on these results, we modeled the heating effect using the low-temperature dependence of specific heat from the Debye model to simulate qubit control with frequency shifts. This model quantitatively reproduces some of the experimental results and makes it possible to simulate qubit gates with temperature-dependent frequency shifts. The simulated control fidelity of the proposed model was consistent with the experimental results, demonstrating that a gate fidelity above $99.9$\% can be achieved in a hot environment. Additionally, our comparison of the two temperature-dependent frequency shift models reveals the condition for a microscopic frequency shift model in which the variance of the frequency shift is limited to a specific range. Furthermore, we investigated gate fidelity using the DRAG method and demonstrated the utility of temperature-dependent simulations.
\end{abstract}


\maketitle

\section{\label{intro} Introduction}

Electrons confined in quantum dots have a spin-1/2 degree of freedom, which can be utilized as qubits with only two levels \cite{lossQuantumComputationQuantum1998}. Such spin qubits are an attractive system for quantum computing due to their long coherence time and the applicability of advanced semiconductor manufacturing technologies for their production \cite{burkardSemiconductorSpinQubits2023}. Recent progress in isotopic purification and control techniques has enabled high control fidelity for both one-qubit gates \cite{takedaOptimizedElectricalControl2018,xueCMOSbasedCryogenicControl2021,noiriFastUniversalQuantum2022,millsTwoqubitSiliconQuantum2022,philipsUniversalControlSixqubit2022,huang2024} and two-qubit gates \cite{noiriFastUniversalQuantum2022,millsTwoqubitSiliconQuantum2022,philipsUniversalControlSixqubit2022}, exceeding the fault-tolerant threshold.

Quantum gate fidelity, however, is limited by intrinsic noise in semiconductor devices, which is known as charge noise \cite{yonedaQuantumdotSpinQubit2018,connors2019}. Experiments using short-channel bulk n-MOSFETs have determined that the source of charge noise in a cryogenic environment is the traps at the semiconductor/oxide interface \cite{inabaDeterminingLowfrequencyNoise2023}. The distributed charge traps model can generate $1/f^{\alpha}$ spectra ($\alpha \approx 1$), measured in many noise spectroscopy experiments \cite{shehataModelingSemiconductorSpin2023}, and reproduce the noise correlations of neighboring qubits \cite{kepaCorrelationsSpinSplitting2023}. While spin qubits do not directly couple to electric fields, spin-orbit coupling exposes them to local electric field fluctuations induced by charge noise. In Si devices, although spin-orbit coupling is weak enough to be negligible, slanting magnetic fields from micromagnets causes spin qubits to be susceptible to charge noise \cite{khaMicromagnetsExposeSpin2015}. Electric dipole spin resonance (EDSR) with slanting magnetic fields \cite{pioro-ladriere2008}, a widely used method for implementing one-qubit gates, uses micromagnets to generate magnetic gradients. Consequently, this introduces charge noise which affects the coherence of spin qubits. For two-qubit gates, charge noise affects the confinement potential, specifically modifying the tunnel coupling related to the strength of the exchange interaction \cite{huangSpinDecoherenceTwoqubit2018}.

It is true that charge noise affects control fidelity through the mechanisms explained above; however, actual fidelity from interleaved randomized benchmarking experiments for several one-qubit gates indicates the existence of another dominant noise source leading to systematic calibration errors, as control fidelity depends on the direction of rotation \cite{yonedaQuantumdotSpinQubit2018}. Subsequent studies have revealed that qubit frequency shifts, which can be as high as about $1$ MHz, are a major obstacle to improving one-qubit gate fidelity \cite{takedaOptimizedElectricalControl2018,zwerver2022,undsethHotterEasierUnexpected2023}. These shifts cause systematic errors in qubit frequency, resulting in direction-dependent control fidelity.

Several studies have simulated spin qubits in silicon quantum dots with the effects of charge noise. For example,  \cite{gungordu2022} added a noise term to the Hamiltonian to create a robust pulse, and  \cite{shehataModelingSemiconductorSpin2023} proposed optimized Full-CI methods for double quantum dots, calculating the strain on the confinement potential from a microscopic charge noise model. However, they did not address the frequency shift mentioned above. The reason may be that the microscopic explanation of frequency shifts is still unknown. Nevertheless, experiments showing the temperature dependence of this phenomenon suggest a relation to control pulse heating effects, and various hypotheses have been presented. For example,  \cite{takedaOptimizedElectricalControl2018} suggested that strain variation modifies the confinement potential, causing frequency shifts. In contrast,  \cite{undsethHotterEasierUnexpected2023} considered many possibilities, including temperature dependence of micromagnets, residual hyperfine interactions, and temperature-dependent electric fields. They reviewed each explanation's validity and conducted additional experiments but could not specify a reasonable microscopic explanation for the frequency shift.  \cite{choiInteractingRandomfieldDipole2024} devised a model using randomly distributed temperature-dependent dipole defects, which can show both positive and negative shifts and non-monotonic temperature dependence. However, as we will demonstrate later, this model cannot explain the high control fidelity in hot environments due to its significant variance.

In this paper, we propose a pulse heating model for temperature-dependent spin qubits simulations of one-qubit gates in silicon quantum dots devices. To reproduce temperature dynamics, we employed the Debye model, considering specific heat and a temperature-dependent cooling power model. We fitted the suggested model to the experimental results of  \cite{undsethHotterEasierUnexpected2023}. To account for qubit frequency fluctuations, we used a model where the frequency shift is a combination of deterministic and stochastic parts. The former is a deterministic function of temperature, fitted to experimental results, while the latter is a temperature-independent charge noise model. This model can reproduce the relationship between base temperature and gate fidelity, showing that a hot environment paradoxically offers higher fidelity gates than a cold environment, as referenced in  \cite{undsethHotterEasierUnexpected2023}. We also explored the conditions for an appropriate model through comparison to the temperature-dependent charge noise model. Finally, we tested the optimization of the Derivative Removal by Adiabatic Gate (DRAG) pulse and demonstrate the utility of temperature-dependent simulations.

The remainder of this paper is organized as follows. Sec. \ref{model} describes models which take into consideration the temperature dependence of silicon spin qubits, including temperature dynamics and charge noise. Sec. \ref{simulation} summarizes methods for simulating one-qubit gates in noisy environments and the configuration of the simulations. Sec. \ref{result} presents the simulation results and their analyses. In Sec. \ref{conclusion}, we summarize our conclusions on the simulation methods.

\section{\label{model} model}
\subsection{\label{sub:heating} Heating dynamics model}
We introduce the heating dynamics model for simulating the heating effect of microwaves in the quantum dot system. Spin qubit devices are ordinarily installed on the dilution refrigerator's mixing chamber to reduce thermal noise and protect qubits from decoherence. This setup keeps the temperature of the device at approximately 10 mK to 1 K \cite{philipsUniversalControlSixqubit2022,huang2024}, but control signals for implementing quantum gates and spin readout cause an unexpected rise in temperature through Joule heating. Experimental results indicate that ac signals, with frequencies ranging from about 10 to 20 GHz, have the greatest effect on temperature \cite{undsethHotterEasierUnexpected2023}. Since only ac signals are needed to implement one-qubit gates, we focus on their effect.

We assume that the heating dynamics are given by
\begin{equation} \label{eq:heating-dynamics}
  \frac{dT(t)}{dt}=\frac{W_\text{Joule} - W_\text{cool}}{C\left(T(t)\right)},
\end{equation}
where $T(t)$ is the device temperature at time $t$, $W_\text{Joule}$ is the Joule heating power of the ac signal, $W_\text{cool}$ is the cooling power of the dilution refrigerator, and $C(t)$ is the heat capacity of the device. Joule heating of the ac signal is assumed \cite{pozar2012}:
\begin{equation} \label{eq:microwave-energy}
  E_\text{Joule}=\int_{t}dt \int_{V} d\mathbf{r}\frac{\sigma(\mathbf{r}) + \omega\epsilon(\mathbf{r})}{2}|E_\text{ac}|^2,
\end{equation} 
where $V$ is the volume of space the fields penetrate. $\sigma(\mathbf{r})$ and $\epsilon(\mathbf{r})$, which are functions of position $\mathbf{r}$, represent conductive and dielectric loss, respectively. From Eq. \eqref{eq:microwave-energy}, we derive a simplified expression of $W_\text{Joule}$:
\begin{equation} \label{eq:microwave-power}
  W_\text{Joule} = (\sigma' + \omega \epsilon')|E_\text{ac}|^2.
\end{equation}
In this model, $\sigma'$ and $\epsilon'$ are determined from experimental results. $W_\text{cool}$ is modeled as 
\begin{equation} \label{eq:cooling-power}
  W_\text{cool} = \alpha_\text{c}T^{\gamma}.
\end{equation}
This model indicates that cooling power has temperature dependence, and its strength is determined by the parameters $\alpha_\text{c}$ and $\gamma$. This is based on the fact that the cooling efficiency of a dilution refrigerator varies depending on the operating temperature \cite{green2015}. In our simulation, we fixed the $\gamma$ at $\gamma=1$ and fitted $\alpha$ to the experimental results. The temperature dependence of heat capacity is modeled based on the Debye model, which predicts the $\propto T^3$ dependency of specific heat at low temperatures:
\begin{equation} \label{eq:heat-capacity}
  C\left(T(t)\right) = k_C T^{3}(t),
\end{equation}
where $k_C$ is a fitting parameter. 

From Eqs. \eqref{eq:heating-dynamics}, \eqref{eq:microwave-power}, \eqref{eq:cooling-power} and  \eqref{eq:heat-capacity}, we can calculate the time variation of temperature. For later fitting and simulation, we present the difference equation form of Eq. \eqref{eq:heating-dynamics}:
\begin{equation} \label{eq:heating-difference-equation}
  T(t+\Delta t)=T(t)+\frac{(\sigma'+\omega\epsilon')|E_\text{ac}|^2 - \alpha_\text{c}T^\gamma(t)}{k_C T^3(t)}\Delta t.
\end{equation}

\begin{figure}[t]
  \centering
  \includegraphics{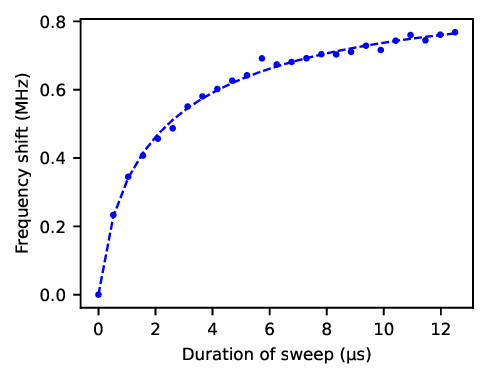}
  \caption{\label{fig:ac_heating_fitting} Measured frequency shifts (circle points) and the fitting curve (dashed line) as a function of the duration of sweep ({\textmu}s). The data was obtained by examining the temperature dependence of the frequency shift and its relationship with ac pulse energy. The proposed model (dashed line) fits the experimental results, demonstrating the frequency shift caused by ac pulse heating. The fitting parameters were calculated based on the theoretical model of ac pulse-induced thermal effects.}
\end{figure}

We fitted this model to the experimental results of  \cite{undsethHotterEasierUnexpected2023}. The fitting procedure is divided into two steps. First, we fitted the function:
\begin{equation} \label{eq:frequency-shift}
  \Delta f(T) = F(T) - F(0).
\end{equation}
to the Q1 frequency shift in Fig. 2 of  \cite{undsethHotterEasierUnexpected2023}. In this model, we defined the function $F(T)$ as
\begin{align}
  F(T) &= \braket{s_+}(T)f_+ + \braket{s_-}(T) f_-,\\
  \braket{s_\pm} &= \pm\frac{1}{\exp\left(-\frac{\mu}{T}\right) + 1} \mp \frac{1}{\exp\left(\frac{\Delta E - \mu}{T}\right) + 1}
\end{align}
where $f_+,f_-,\mu,$ and $\Delta E$ are fitting parameters. This model is equivalent to the two dipole defects model similar to that in  \cite{choiInteractingRandomfieldDipole2024}. Second, we fitted the parameters in difference equation  \eqref{eq:heating-difference-equation} into the Q1 results of Fig. 3 in  \cite{undsethHotterEasierUnexpected2023}. Fig. \ref{fig:ac_heating_fitting} shows the fitting result. The dashed line indicates the fitting result for Eq. \eqref{eq:heating-difference-equation}, which is consistent with experimental results. Note that while frequency $\omega$ was set to $16$ MHz as an off-resonant pulse within the fitting step, we set $\omega$ to $15.6$ MHz as the resonant pulse. Further details and additional information about parameter fitting are described in Appendix  \ref{adx:fitting-detail}.

\subsection{\label{sub:charge-noise} Charge noise model}
To approximate the noise which has $1/f$-like power spectral density(PSD), we adopted a random distributed two-level fluctuators (TLFs) model \cite{dutta1981}. TLFs-generated noise is known as random telegraph noise (RTN) and exhibits two states (referred to as $0$ and $1$) with characteristic switching times. Its PSD is expressed by
\begin{equation} \label{eq:rtn-psd}
  P_\text{RTN}(f)=\frac{2A^2}{(\tau_0 + \tau_1)\left\{\left(\frac{1}{\tau_0} + \frac{1}{\tau_1}\right)^2 + (2\pi f)^2\right\}},
\end{equation}
where $A$ is the peak-to-peak amplitude of noise, $\tau_0(\tau_1)$ are the switching times of $0$ to $1$($1$ to $0$), respectively, and $f$ denotes frequency \cite{machlup1954}. When different TLFs have varying switching times, their superposition results in a $1/f$-like PSD noise.

The noise model varies depending on the type of TLFs.  \cite{rojas-ariasSpatialNoiseCorrelations2023} used a charge trap model with screening due to metal structure. They approximate screening by introducing an image charge $-\delta q_i$ to each trapped charge $\delta q_i$, which effectively predicts noise correlations. \cite{shehataModelingSemiconductorSpin2023} explored a double well potential model where charges switch between two potential wells. We chose the former model for its simplicity, but we believe similar outcomes can be obtained with the latter model.

In the screening model, the $i$th trapped charge $\delta q_i$ at position $\mathbf{r}_i$ generates an electric field:
\begin{equation} \label{eq:ith-electric-field}
  \delta \mathbf{E}_i = \frac{1}{4\pi \epsilon_\text{r} \epsilon_0} \left\{ \frac{\mathbf{r} - \mathbf{r}_i}{|\mathbf{r} - \mathbf{r}_i|^3}\delta q_i -\frac{\mathbf{r} - \mathbf{r}'_i}{|\mathbf{r} - \mathbf{r}'_i|^3}\delta q_i \right\},
\end{equation}
at $\mathbf{r}$. Here, $\mathbf{r}'_i$ is the position of the image charge, $\epsilon_\text{r}$ is the effective dielectric constant of the layer, and $\epsilon_0$ is the permittivity of vacuum. We defined the plane quantum dot formed as $z=0$, with charges distributed at $z=50$ nm, and placed the image charge at $z=54$ nm, following the structure of  \cite{rojas-ariasSpatialNoiseCorrelations2023}. The charge traps were assumed to be distributed within a square area of $300$ nm on each side, centered on the quantum dot.

The frequency shift $\delta f$ due to the electric field $\delta \mathbf{E}=(\delta E_x, \delta E_y, \delta E_z)$ can be calculated using the following equation \cite{rojas-ariasSpatialNoiseCorrelations2023}:
\begin{equation}
  \delta f = \frac{g\mu_B}{h}\frac{e}{m\omega_\text{c}^2}\left(\frac{\partial B}{\partial x}\delta E_x + \frac{\partial B}{\partial z}\delta E_y\right)
\end{equation}
where $\hbar$ is the reduced Planck's constant, $\hbar \omega_\text{c}$ is the confinement energy of the electron, $B$ is the $z$ component of the magnetic field at the quantum dot's position, $g$ is the $g$-factor, $\mu_B$ is the Bohr magneton, $e$ is the electron charge, and $m$ is the effective mass of the electron. We neglected the $y$ component of the electric field due to the strong confinement along the $z$-direction, where the electron's position change is minimal. The parameter values used in this paper are shown in Table \ref{table:parameters}.

\begin{table}[b]
  \caption{\label{table:parameters} Parameters used for the calculation during a simulation. These values were determined based on  \cite{rojas-ariasSpatialNoiseCorrelations2023, shehataModelingSemiconductorSpin2023}. Note that $m_0$ denotes the mass of an electron and the confinement was assumed to be parabolic.}
  \begin{ruledtabular}
    \begin{tabular}{cc}
      Parameter& Value\\
      \colrule
      $\partial B/\partial x$& $0.15$ mT/nm\\
      $\partial B/\partial y$& $0.01$ mT/nm\\
      $g$ & $2$\\
      $\epsilon_\text{r}$ & $13$\\
      $m$ & $0.19m_0$ kg\\
      $\hbar \omega_\text{c}$ & $3$ meV\\
    \end{tabular}
  \end{ruledtabular}
\end{table}

To simulate the temporal change in charge noise, the state transition probability must be calculated for each charge trap. In the limit where $\tau_{0(1)}\ll \Delta t$, these probabilities $p$ are given by
\begin{equation} \label{eq:transition-probability}
  p_{0\rightarrow 1 (1\rightarrow 0)} = \tau_{0(1)}\Delta t.
\end{equation}
Using this probability, we can simulate the stochastic dynamics of charge noise over time.

\subsection{\label{sub:td-charge-noise} Temperature-dependent charge noise model}
The frequency shift model proposed by  \cite{choiInteractingRandomfieldDipole2024} incorporates the temperature dependence of charge trap fluctuations. In their non-interacting model, dipole defects with different energy levels between charge states lead to a definite expectation value of frequency shift $\delta f$ at $T=0$, converging to $\delta f = 0$ ad $T\rightarrow \infty$. When multiple dipole defects are appropriately distributed, they induce non-monotonic behavior in the frequency shift, where the frequency shift of each defect can be both positive and negative.

We adapt this concept using the screening model and express the temperature dependence of switching time as \cite{dutta1981,connors2019}:
\begin{align}
  \tau_0(T) &= \tau_0 \exp\left(\frac{E_i}{T}\right),\\
  \tau_1(T) &= \tau_0 \exp\left(\frac{E_i - \Delta E_i}{T}\right),
\end{align}
where $i$th $E_i$ and $\Delta E_i$ for each $i$ are extracted from a uniform distribution with a range of $100$ to $1,000$ $\text{K}^{-1}$. $\tau_0$ is extracted from a log-uniform distribution with a range of $10^{3}$ ns to $10^9$ ns. At low temperatures, these switching times can become exceedingly large, so we set an upper limit of $10^{15}$ ns for numerical stability. This does not affect the results within the simulated time range.

\begin{figure}[t]
  \centering
  \includegraphics{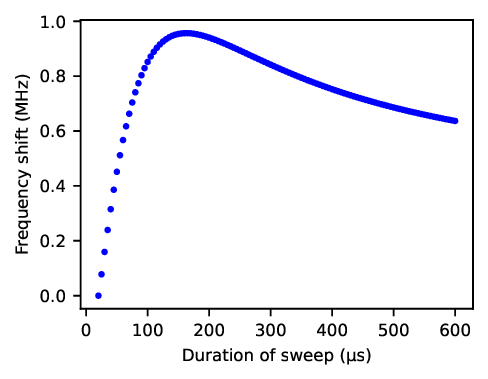}
  \caption{\label{fig:nonmonotonic-frequency-shift} Reproduction of temperature dependence of frequency shift based on a temperature-dependent charge noise model. This model successfully demonstrates the non-monotonic behavior of the frequency shift as the temperature increases, indicating a peak frequency shift at around 150 mK followed by a decrease at higher temperatures.}
\end{figure}

This temperature-dependent model can exhibit non-monotonic variation (Fig. \ref{fig:nonmonotonic-frequency-shift}). Note that the magnetic gradients used in simulations are larger than those listed in Table \ref{table:parameters}, as the previous values were insufficient for achieving realistic frequency shifts.

\section{\label{simulation} temperature-dependent simulation}
\subsection{Unitary evolution}
In the simulation, we assumed the implementation of an X gate using EDSR. The Hamiltonian in the rotating frame is described by:
\begin{equation} \label{eq:hamiltonian}
  H(t) = \frac{1}{2}f_x(t)\sigma_x + \frac{1}{2}\Delta f_\text{noise}(t)\sigma_z,
\end{equation}
where $\sigma_x$ and $\sigma_z$ are Pauli operators, $f_x(t)$ is a time-dependent parameter controlled electrically as per the EDSR method \cite{pioro-ladriere2008}, and $\Delta f_\text{noise}(t)$ represents stochastic deviations from a calibrated Larmor frequency. For a gate time $t_\text{g}$, the time evolution is divided into N steps $\Delta t_1,\Delta t_2,...,\Delta t_N$. The time evolution during $i$th time step can be approximated by the following unitary operator:
\begin{equation} \label{eq:time-step-unitary}
  U_i \approx \exp\left\{-i \frac{2\pi H(t_i)}{\hbar}\Delta t\right\},
\end{equation}
where $t_i =\sum_{j=1}^{i}\Delta t_j$. The resulting quantum gate is expressed as:
\begin{equation} \label{eq:unitary-evolution}
  U_\text{g} \approx U_N U_{N-1}\ldots U_1.
\end{equation}
To implement an X gate effectively, $f_x(t)$ must be controlled so that the gate fidelity,
\begin{equation} \label{eq:gate-fidelelity}
  F = \frac{1}{2}\text{Tr}\left[U_\text{target}^\dagger U_\text{g} \right]
\end{equation}
approaches 1. Here, $\text{Tr}$ denotes the trace of an operator, and $U_\text{target}$ is set to $\sigma_x$.

\begin{table*}[t]
  \caption{\label{table:compare-x-fideliteis} X gate fidelities from 10 simulations based on Model 1 and Model 2. Each fidelity value is the mean of 40,000 gate executions, with the temperature reset every 400 executions. The values shown in the bottom row are the mean fidelity of 10 simulations. Note that the values in the table are rounded to four decimal places.}
  \begin{ruledtabular}
  \begin{tabular}{ccccccccccc}
   $T_0$(mK)& &20&100&200&300& &20&100&200&300\\
   \colrule
   &Model 1&0.9957&0.9865&0.9909&0.9926&Model 2&0.9879&0.9959&0.9997&0.9997\\
   & &0.9950&0.9949&0.9747&0.9955& &0.9909&0.9959&0.9997&0.9997\\
   & &0.9959&0.9948&0.9661&0.9923& &0.9916&0.9959&0.9997&0.9997\\
   & &0.9953&0.9951&0.9930&0.9867& &0.9907&0.9959&0.9997&0.9997\\
   & &0.9948&0.9792&0.9955&0.9929& &0.9908&0.9959&0.9997&0.9997\\
   & &0.9916&0.9952&0.9803&0.9899& &0.9907&0.9959&0.9997&0.9997\\
   & &0.9943&0.9930&0.9948&0.9939& &0.9907&0.9959&0.9997&0.9997\\
   & &0.9956&0.9948&0.9951&0.9899& &0.9907&0.9959&0.9997&0.9997\\
   & &0.9947&0.9898&0.9885&0.9951& &0.9907&0.9959&0.9997&0.9997\\
   & &0.9957&0.9922&0.9908&0.9832& &0.9907&0.9959&0.9997&0.9997\\
   \colrule
   &Mean&0.9949&0.9916&0.9870&0.9912& &0.9905&0.9959&0.9997&0.9997
  \end{tabular}
  \end{ruledtabular}
\end{table*}

\subsection{Gaussian pulse}
We employed a Gaussian pulse for implementing an X gate, defined as:
\begin{equation} \label{eq:gaussian-pulse}
  f(t) = \gamma_0 \exp\left\{-\frac{\left(t-t_\text{g}/2\right)^2}{2\sigma^2}\right\}
\end{equation}
where $\sigma$ is the width of the pulse set to $t_\text{g}/6$. The amplitude $\gamma_0$ is determined by:
\begin{equation} \label{eq:amplitude}
  \gamma_0 = \frac{1}{2\sqrt{2\pi}\sigma\:\text{erf}\left(\frac{t_\text{g}}{2\sqrt{2}\sigma}\right)},
\end{equation}
where $\text{erf}(\cdot)$ denotes the error function.

While Eq. \eqref{eq:gaussian-pulse} represents the amplitude in terms of frequency, the heating simulation requires the microwave amplitude $E_\text{pulse}$ to be in volts. Thus, we converted the frequency $f$(MHz) to voltage $E_\text{pulse}$(V) using:
\begin{equation}
  E_\text{pulse} = \frac{f}{1.3}\times 0.2.
\end{equation}
This conversion is based on the experimental result, i.e., $1.3$ MHz Rabi frequency with a $0.2$ V microwave amplitude \cite{undsethHotterEasierUnexpected2023}. Note that we assumed linearity between the microwave amplitude and Rabi frequency holds within the range considered in this paper, but this may not hold in an actual device \cite{philipsUniversalControlSixqubit2022}.

\subsection{Simulation method}
At the beginning of our simulation, we conducted pseudo-calibration to account for variations in the Larmor frequency due to changes in the base temperature. In real experiments, this calibration is performed through specific measurements, such as fitting to a chevron pattern. In this paper, however, we approximate this process by collecting frequency changes at specific time intervals during a given calibration period and calculating the mean frequency change. While this procedure does not exactly replicate real experimental conditions, it provides a reasonable estimate of the values expected from the calibration step. The measurement duration and interval were set to $10^{9}$ ns and $50$ ns, respectively.

In the simulation, we collected the fidelity of $40,000$ gates per execution. The gate time ($t_\text{g}$) was set to $100$ ns, and the cooling time between each gate ($t_\text{cool}$) was set to $50$ ns. The Hamiltonian and temperature were updated at a time step of 0.1 ns, while the charge noise states were updated every 10 ns due to the relatively slow transition of trapped charge states.

To mimic the number of consecutive gate executions of randomized benchmarking, we reset the temperature to the base level after every $400$ gate executions. Following the temperature reset, the charge noise states were updated using Eq. \eqref{eq:transition-probability} with $\Delta t$ set to $10,000$ ns. Although this $\Delta t$ exceeds the typical approximation limits, we restricted the transition probability to a maximum of 0.5 to ensure the generation of valid random states. We believe that this incorrectness does not significantly affect the validity of the primary simulation results. Note that to keep the mean frequency shift consistent, the temperature-dependent charge trap was forced to its initial state upon temperature reset, regardless of the reset time.

\section{\label{result} simulation results}
\subsection{Base temperature dependence of gate fidelity}

We compared two temperature-dependent frequency shift models:

\begin{enumerate}[align=left, left=5pt]
  \item[Model 1] Temperature-dependent charge noise
  
    This model represents the frequency shift as electric field fluctuations resulting from 30 charge traps with temperature-dependent switching times, as explained in Sec. \ref{sub:td-charge-noise}. To consider the effect of low-frequency noise, 20 TLFs with constant switching times ranging from $10^9$ to $10^{15}$ ns were used.
  \item[Model 2] Decisive frequency shift + charge noise
  
  This model represents the frequency shift as a decisive function of temperature fitted through actual experimental results, as explained in Sec. \ref{sub:heating}. Additionally, temperature-independent charge was incorporated as a random frequency noise source, as explained in Sec. \ref{sub:charge-noise}. The number of traps was set to 50.
\end{enumerate}

Model 1 is a microscopic model designed to reproduce the frequency shift, whereas Model 2 only fits to the experimental data. Therefore, the simulation results of Model 2 are expected to be consistent with the actual results. By comparing these models, we can assess the validity of Model 1. Moreover, Model 2 will be useful for simulations under various conditions, such as when using different pulse amplitudes or shapes.

\begin{figure}[b]
  \centering
  \includegraphics{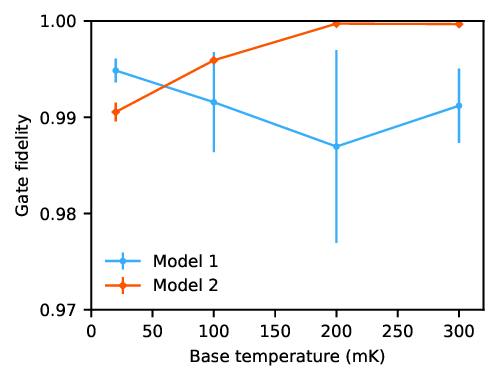}
  \caption{\label{fig:temperature-dependence-simulation} Relationship between base temperature and gate fidelity to Model 1 (blue) and Model 2 (orange). The error bars represent one standard deviation of the 10 mean gate fidelities. While Model 1 did not align with the experimental results, which showed improved gate fidelity in a hotter environment, Model 2 successfully replicated the results.}
\end{figure}

For each model, we simulated X gate execution with the configuration described in Sec. \ref{simulation}, using base temperatures of $T_0=20,100,200,300$ mK. The X gate fidelities are shown in Table \ref{table:compare-x-fideliteis}, representing the results of 10 simulations for each model and temperature. The mean fidelity changes are plotted in Fig. \ref{fig:temperature-dependence-simulation}. 

In Model 1, the highest fidelity was obtained at the lowest temperature, which is inconsistent with the experimental results in  \cite{undsethHotterEasierUnexpected2023}. In contrast, the temperature dependence of the gate fidelities in Model 2 is more reasonable as the X gate fidelity in the hotter environments($T_0=200,300$ mK) was higher than the colder environment($T_0=20,100$ mK).

The inconsistency of Model 1 stems from the large variance of the temperature-dependent charge noise model. To reproduce the frequency shift in the actual device, the effect of each charge trap must be greatly estimated. As a result, while the mean value of the frequency shift is reproduced from these traps, each realization of frequency shift during the consecutive gate executions varies widely and the deviation from the mean value increases with rising temperature. This leads to the high variance in fidelity, as shown by the long error bar in Fig. \ref{fig:temperature-dependence-simulation}.

\begin{figure}[b]
  \centering
  \includegraphics{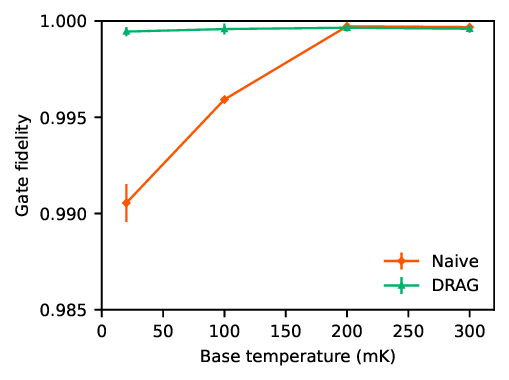}
  \caption{\label{fig:drag-simulation} Relationship between base temperature and gate fidelity for the DRAG pulse (green) compared with previous results using a naive Gaussian pulse (orange) in Model 2. The error bars represent one standard deviation of the 10 mean gate fidelities. While the DRAG pulse improved fidelity at low temperatures, it was less effective at higher temperatures.}
\end{figure}

\subsection{DRAG pulse simulation}

\begin{table}[t]
  \caption{\label{table:drag-fidelities} Adjusted $\alpha$ and X gate fidelities to each base temperature. Each fidelity value is the mean of 40,000 gate executions, with the temperature reset every 400 executions. Note that the values of fidelities in the table are rounded to four decimal places.}
  \begin{ruledtabular}
    \begin{tabular}{ccc}
      $T_0$(mK)&$\alpha$&Gate Fidelity\\
      \colrule
      20&$-0.1444$&0.9995\\
      100&$-0.0778$&0.9996\\
      200&0.0061&0.9997\\
      300&0.0203&0.9996
    \end{tabular}
  \end{ruledtabular}
\end{table}

To demonstrate the utility of temperature-dependent simulation, we optimized the DRAG pulse parameter using the fitted frequency shift model of Eq. \eqref{eq:frequency-shift}, and evaluated the performance in Model 2. DRAG was originally proposed to inhibit leakage, a type of noise observed in qubits with additional levels beyond a two-level system, such as superconducting or ion trap qubits. This is achieved through an additional control pulse comprised of the time-derivative of the original pulse \cite{motzoi2009}. This technique is also effective for canceling frequency shifts \cite{takedaOptimizedElectricalControl2018}.

To implement DRAG, we modify the Hamiltonian in Eq. \eqref{eq:hamiltonian} as follows:
\begin{equation} \label{eq:drag-hamiltonian}
  H(t) = \frac{1}{2}f_x(t)\sigma_x + \frac{1}{2}f_\text{DRAG}(t)\sigma_y + \frac{1}{2}f_\text{noise}(t)\sigma_z.
\end{equation}
Here, we added the second term as a second control pulse. The coefficient $f_\text{DRAG}(t)$ can be calculated from $f_x$ as:
\begin{equation} \label{eq:drag-pulse}
  f_\text{DRAG}(t) = -\frac{\alpha \gamma_0(t-t_\text{g}/2)}{\sigma}\exp\left\{-\frac{\left(t-t_\text{g}/2\right)^2}{2\sigma^2}\right\}
\end{equation}
where $\alpha$ is the parameter that needs to be adjusted. Note that we used Eq. \eqref{eq:gaussian-pulse} to replace $f_x(t)$ with a Gaussian pulse. The energy caused by the DRAG pulse is assumed to be the root mean square of each pulse.

Table \ref{table:drag-fidelities} shows the results of adjusting the parameter $\alpha$ and the mean fidelities obtained from 10 sets of simulations in Model 2 for each base temperature. The relationship between base temperature and fidelity is plotted in comparison with the results of the naive Gaussian pulse in Fig. \ref{fig:drag-simulation}. At low temperatures ($T_0=20,100$ mK), the fidelity was improved by the DRAG pulse, as reported in  \cite{takedaOptimizedElectricalControl2018}. However, at higher temperatures ($T_0=200,300$ mK), the fidelity decreased compared to the naive pulse, and the variance of fidelity widened, suggesting that the DRAG pulse is more susceptible to noise and less effective in high-temperature environments. This may result from the heating effect of the additional pulse outweighing the effect of increased fidelity and suggests that a sophisticated pulse shaping method will be necessary to improve fidelity in hotter environments.

\section{\label{conclusion} conclusion}
We implemented temperature-dependent simulations of spin qubits in silicon quantum dots. Our work involved modeling the relationship between microwave amplitude and temperature as well as temperature dependence of Larmor frequency shifts. By comparing our simulation results with the experimental gate fidelity from real devices, we found that the temperature-dependent charge noise model is not appropriate for explaining frequency shifts due to its large variance. Conversely, the frequency shift model based on experimental fitting combined with temperature-independent charge noise successfully reproduced the experimental observation that a hotter environment allows for more precise gate control. Although this model does not provide a microscopic source for frequency shifts, it highlights the conditions that a genuine model must satisfy. Specifically, any microscopic model explaining Larmor frequency shifts should avoid having such large variance that it cancels out the calibration intended to obtain the mean frequency shift.

Furthermore, we emphasize the importance of temperature-dependent simulations as conventional simulations typically consider only charge noise and ignore the effects of heating \cite{shehataModelingSemiconductorSpin2023,jnane2024}. While such simulations can be useful to some extent, they have limitations when it comes to optimizing qubit control. We demonstrated the utility of temperature-dependent simulations through the optimization of the DRAG pulse. We believe that more sophisticated temperature dependence models will provide a beneficial platform for simulating and optimizing qubit control of spin qubits in quantum dots.

\begin{figure}[b]
  \centering
  \includegraphics{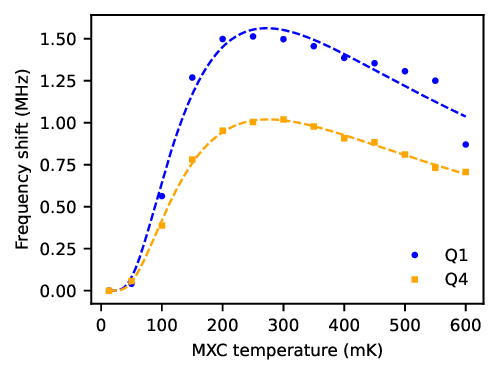}
  \caption{\label{fig:frequency-shift-fit} Fitting results of Eq. \eqref{eq:frequency-shift} for the Q1 (blue) and Q4 (orange) from  \cite{undsethHotterEasierUnexpected2023}. Both fitting curves (dashed lines) fit the experimental data well (circles for Q1 and squares for Q4). Only two parameters $f_+$ and $f_-$ differ between the curves while $\mu$ and $\Delta E$ are obtained from the fit for Q1 and are common to both curves.}
\end{figure}


\appendix
\section{\label{adx:fitting-detail} Detail explanation of fitting}

\begin{figure}[t]
  \centering
  \includegraphics{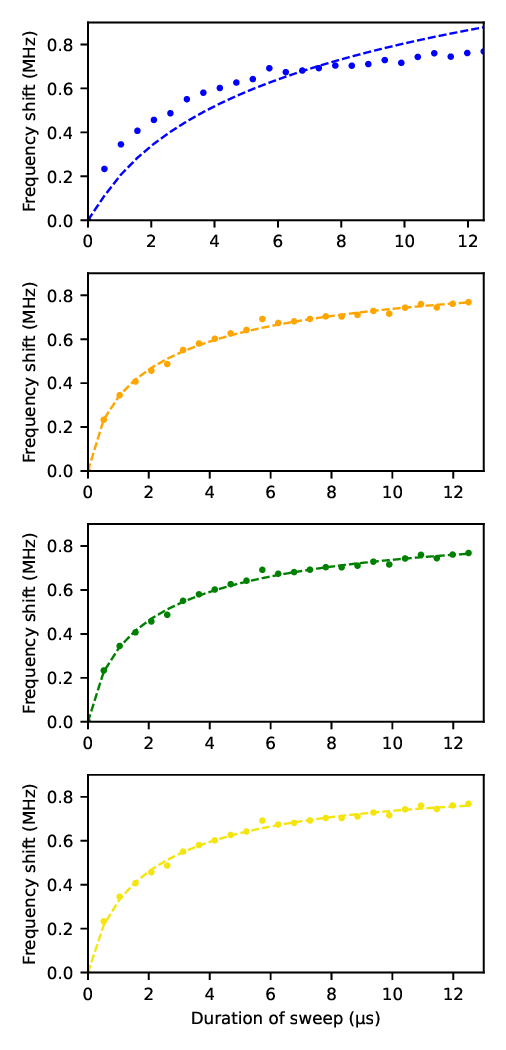}
  \caption{\label{fig:compare-heating-fitting} Fitting curves (dashed lines) of heating model for experimental data (circles) to different $\gamma$. These curves fit the experimental data well except for $\gamma=0$. This demonstrates the importance of incorporating the temperature dependency of cooling power.}
\end{figure}

As described in the main text, we fitted Eq. \eqref{eq:frequency-shift} to the experimental results of frequency shift. This process was conducted using least-squares fitting. Fig. \ref{fig:frequency-shift-fit} shows the experimental data and fitting results. The blue dashed line represents the fitting curve to the Q1 frequency shifts in  \cite{undsethHotterEasierUnexpected2023}. Although there are deviations between the data points and fitting curve in the high temperature region, the curve captures the general trend of temperature dependence. It is worth noting that the fitting curve for the Q4 result (orange points and curve in Fig. \ref{fig:frequency-shift-fit}) in  \cite{undsethHotterEasierUnexpected2023} was done by only adjusting only the parameters $f_+$ and $f_-$, and fixing the parameters $\mu$ and $\Delta E$ obtained from the preceding fitting. The two quantum dots Q1 and Q4 are on the same device, suggesting that frequency shifts have a strong correlation with neighboring qubits and may share the same intrinsic source. This observation supports the notion that the temperature-dependent charge noise model cannot fully explain the frequency shift, as peculiar arrangements of charge traps would be necessary to reproduce this strong correlation.

Following the fitting to the temperature-dependence of frequency shifts, we fitted the parameters in Eq. \eqref{eq:heating-difference-equation} to the relationship between the input energy and frequency shift of Q1 in  \cite{undsethHotterEasierUnexpected2023}. While we chose $\gamma = 1$ in the main text, here, we compare the fitting results with $\gamma=0,0.5,1,2$. As shown in Fig. \ref{fig:compare-heating-fitting}, fitting curves for each value of $gamma$ successfully reproduce the experimental results except for $\gamma=0$, demonstrating the validity of considering the temperature dependency of cooling power.

\begin{figure}[t]
  \centering
  \includegraphics{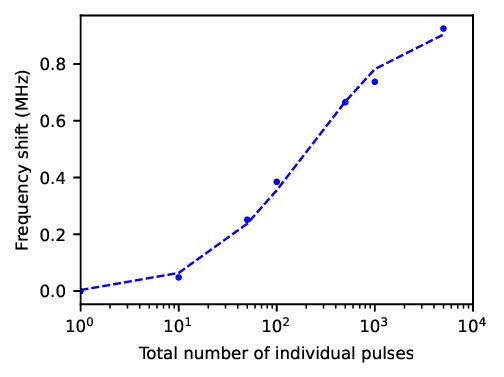}
  \caption{\label{fig:dc-heating-fitting} Result of fitting heating model for dc pulse heating due to the B5 gate in  \cite{undsethHotterEasierUnexpected2023}. Fitting lines capture the non-linear change in the frequency shift. The horizontal axis represents the logarithmic scale.}
\end{figure}

We also fitted the parameters in Eq. \eqref{eq:heating-difference-equation} (with $\omega=0$) to the frequency shift induced by a dc pulse. This fitting was conducted using the experimental results for the B5 gate in  \cite{undsethHotterEasierUnexpected2023} (with $\gamma$ set to 1). As shown in Fig. \ref{fig:dc-heating-fitting}, the fitting curve is consistent with the data points from the experiment. Note that the parameter values obtained from fitting to ac and dc heating are not consistent. The discrepancy likely arises from differences between ac and dc pulse sources. It is evident from $ \eqref{eq:microwave-energy}$ that the heating effect varies with the heated position, resulting in different fitting parameters.

\begin{table}[b]
  \caption{\label{table:compare-temperature} Comparison of attained temperatures for the Gaussian and square pulse at each base temperature.}
  \begin{ruledtabular}
    \begin{tabular}{ccc}
      $T_0$(mK)&Gaussian(mK)&Square(mK)\\
      \colrule
      20&483.92&410.86\\
      100&484.09&411.11\\
      200&486.77&415.07\\
      300& 498.46&432.41\\
    \end{tabular}
  \end{ruledtabular}
\end{table}

\begin{figure}[t]
  \centering
  \includegraphics{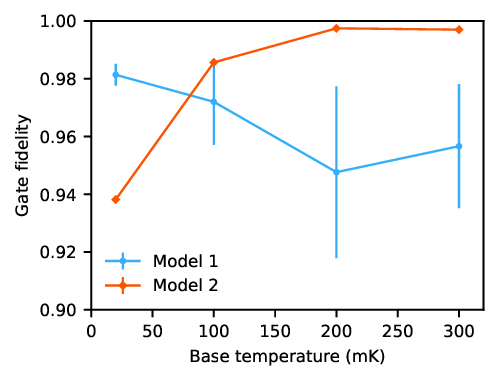}
  \caption{\label{fig:compare-low-temperature} Relationship between base temperature and gate fidelity to Model 1 (blue) and Model 2 (orange) when the heating power is lower than that in the main text. The error bars represent one standard deviation of the 10 mean gate fidelities. The overall trends in gate fidelity with temperature changes for each model are similar to those in Fig. \ref{fig:temperature-dependence-simulation} in the main text.}
\end{figure}

\section{\label{adx:additional-simulation} Additional simulation results}
\subsection{\label{sub:weak-heating} Weak heating}

To investigate scenario where the attained temperature is lower than in the scenario examined in the main text (gate time is $t_\text{g} = 100$ and reset interval is 400 gate executions), we ran simulations with a gate time of $t_\text{g} = 400$ ns and reset the temperature every 200 gate executions. Extending the gate time reduces the amplitude of the Gaussian pulse, which in turn lowers the final temperature. Fig. \ref{fig:compare-low-temperature} shows a comparison of simulations based on Model 1 and Model 2. The overall trends in gate fidelity with temperature changes for each model are similar to those in Fig. \ref{table:compare-x-fideliteis} in the main text. This indicates that the significant variance in frequency shift leads to inconsistent simulations, even at relatively low temperatures.

Furthermore, compared to the results in Fig. \ref{fig:temperature-dependence-simulation}, gate fidelity declines with the longer gate time for all base temperatures. The reason is that the longer the time affected by the Larmor frequency variation, the greater the deviation of the trajectory depicted by the qubit's state from the direct transition.

\subsection{\label{sub:square} Square pulse}

\begin{figure}[t!]
  \centering
  \includegraphics{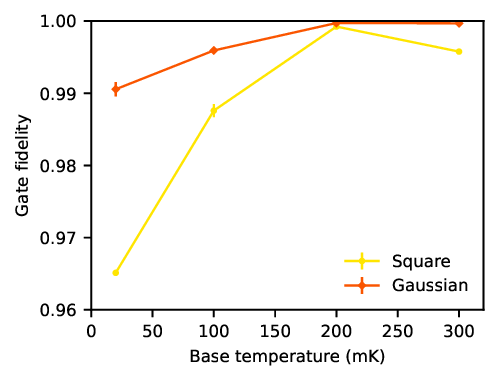}
  \caption{\label{fig:square-pulse-simulation} Relationship between base temperature and gate fidelity to the square pulse (yellow) and Gaussian pulse (orange) in Model 2. The error bars represent one standard deviation of the mean gate fidelities from 10 simulations. While the overall trend of base temperature dependency is similar for both pulses, the gate fidelity obtained from Gaussian pulse is higher than that of the square pulse.}
\end{figure}

In the main text, we used a Gaussian pulse. Here, we evaluate the performance of square pulse, which has a constant amplitude and represents an analytic solution. We simulated the implementation of X gates using a square pulse in Model 2. As shown in Fig. \ref{fig:square-pulse-simulation}, the gate fidelity of the square pulse is lower than that of the Gaussian pulse. This result demonstrates the robustness of the Gaussian pulse against fluctuations in Larmor frequency. On the contrary, the heating effect of the square pulse is lower than that of the Gaussian pulse, as shown in Table \ref{table:compare-temperature}. However, this does not contribute to improved gate fidelity, which is consistent with the observation that frequency shift and temperature are less likely to change in high-temperature environments.

\newpage

\bibliography{bib_hm_manu}

\end{document}